\documentclass[preprint,showpacs,showkeys,preprintnumbers,amsmath,amssymb]{revtex4}
\usepackage{graphicx}% Include figure files
\usepackage{dcolumn}% Align table columns on decimal point
\usepackage{bm}% bold math

\def\ut#1{\mathop{\vtop{\ialign{##\crcr
     $\hfil\displaystyle{#1}\hfil$\crcr\noalign
     {\kern1pt\nointerlineskip}\hbox{$\hfil\sim\hfil$}\crcr
     \noalign{\kern1pt}}}}}

\begin{document}

\preprint{}

\title{Solar system
constraints on R$^n$ gravity}% Force line breaks
%with \\

\author{A.F. Zakharov}
\email{zakharov@itep.ru}
 \affiliation{National Astronomical Observatories of
Chinese Academy of Sciences, Beijing 100012, China}
 \altaffiliation[Also at ]{Institute of
Theoretical and Experimental Physics, Moscow, 117259, Russia;\\
  Joint Institute for Nuclear Research, Dubna,   Russia
%Lines break
%automatically or can be forced with \\
}
\author{A.A. Nucita}%
 \email{achille.nucita@le.infn.it}
\affiliation{Department of Physics and {\it INFN}, University of
Lecce, CP 193, I-73100 Lecce, Italy}\emph{}
\author{F. De Paolis}%
 \email{Francesco.DePaolis@le.infn.it}
\affiliation{Department of Physics and {\it INFN}, University of
Lecce, CP 193, I-73100 Lecce, Italy}
\author{G. Ingrosso}
\email{Gabriele.Ingrosso@le.infn.it}
\affiliation{Department of Physics and {\it INFN}, University of
Lecce, CP 193, I-73100 Lecce, Italy}

\date{\today}% It is always \today, today,

\begin{abstract}
Recently, gravitational microlensing has been investigated in the
framework of the weak field limit of fourth order gravity theory.
However, solar system data (i.e. planetary periods and light
bending) can be used to put strong constraints on the parameters
of this class of  gravity theories. We find that these parameters
must be very close to those corresponding to the Newtonian limit
of the theory.
\end{abstract}

%Use showkeys class option if keyword
\maketitle

\section{\label{intro} Introduction }%\lowercase{via} \textbackslash\textbackslash}

%\section{\label{sec:level1}First-level heading:\protect\\ The line
%break was forced \lowercase{via} \textbackslash\textbackslash}

Since many years different alternative approaches to gravity have
been proposed in the literature such as MOND
\cite{Milgrom_1983,Milgrom_2003}, scalar-tensor \cite{Brans_1961},
conformal  \cite{Pervushin}, Yukawa-like corrected gravity
theories \cite{Fischbach_1986,Fischbach_1992,Hoyle_2001}, and so
on (see papers \cite{Will_2006} for reviews). Very recently, it
has been proposed \cite{Cap_05,Cap_06_04,Cap_06_03}, in the
framework of higher order theories of gravity -- also referred to
as $f(R)$ theories -- a modification of the gravity action with
the form

\begin{equation}
{\cal A}=\int d^4 x \sqrt{-g}[f(R) + {\cal L}_m],
\end{equation}
where $f(R)$ is a generic function of the Ricci scalar curvature
and ${\cal L}_m$ is the standard matter Lagrangian. For example,
if $f(R)=R+2\Lambda$ the theory coincides with General Relativity
(GR) with the $\Lambda$ term. In particular,
\citet{Cap_06_04,Cap_06_03} considered  power law function $f(R)$
theories of the form $f(R)=f_0R^n$. As a result, in the weak field
limit \cite{footnote0}, the gravitational potential is found to be
\cite{Cap_06_04,Cap_06_03}:
\begin{equation}
\Phi(r)=-\frac{Gm}{2r}
\left[1+\left(\frac{r}{r_c}\right)^{\beta}\right]~,
\label{potential}
\end{equation}
where
\begin{equation}
\beta=\frac{12n^2-7n -1
-\sqrt{36n^4+12n^3-83n^2+50n+1}}{6n^2-4n+2}~.
\end{equation}
\begin{figure}[t!]
\begin{center}
\includegraphics[width=0.70\textwidth]{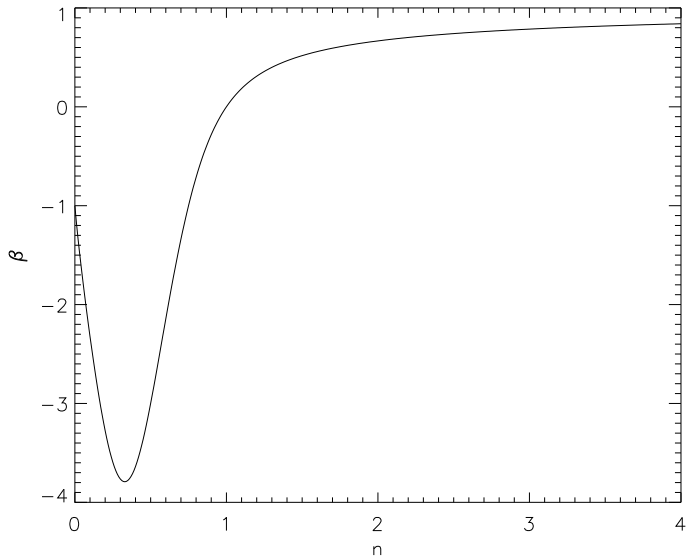}
\end{center}
%\vspace{-1.3cm}
\caption{The parameter $\beta$ as a function of $n$ for fourth order gravity.} %\vspace{-0.5cm}
\label{beta_dependence}
\end{figure}
The dependence of the $\beta$ parameter on the $n$ power is shown
in Fig.~\ref{beta_dependence}. Of course, for $n\rightarrow \infty
$ it follows $\beta\rightarrow 1$, while for $n=1$ the parameter
$\beta$ reduces to zero and the Newtonian gravitational field is
recovered. On the other hand, while $\beta$ is a universal
parameter, $r_c$ in principle is an arbitrary parameter, depending
on the considered system and its typical scale. Consider for
example the Sun as the source of the gravitational field and the
Earth as the test particle. Since Earth velocity is $\simeq 30$ km
s$^{-1}$, it has been found that the parameter $r_c$ varies in the
range $\simeq 1-10^4$ AU.
%The authors fix $r_c$ to be $r_c=Gm/v^2$, $m$ being the solar mass
%and $v \sim$~30~km s$^{-1}$, in order to have no deviation from
%Newtonian potential for a circular Earth orbit with $r\simeq 1$
%A.U., meaning that $r_c$ has to be $1$ A.U.
Once $r_c$ and $\beta$ has been fixed, \citet{Cap_06_04} used them
to study deviations from the standard Paczynski light curve for
gravitational microlensing \cite{Pacz_1986} and claimed that the
implied deviation can be measured~\cite{footnote}. It is clear
that for gravitational microlensing one could detect observational
differences between GR and an alternative theory (the fourth order
gravity in particular), so that one should have different
potentials at the scale $R_E$ (the Einstein radius) of the
gravitational microlensing . For the Galactic microlensing case
$R_E$ is about 1 AU. This is a reason why the authors
\cite{Cap_06_04} have selected $r_c$ at a level of astronomical
units to obtain observable signatures for non-vanishing $\beta$.
 The aim of the
present paper is to show that solar system data (light bending and
planetary periods) put extremely strong constraints on both $r_c$
and $\beta$ parameters making this alternative theory of gravity
not so attracting.

\section{\label{Solar} Solar System Constraints} %\textbackslash\textbackslash}

\subsection{\label{sec:Solar_light bending}Light Bending Constraints}

 As stated above, we now discuss some observational
consequences of the fourth order gravity, depending on the choice
of the parameters $\beta$ and $r_c$.

\begin{figure}[t!]
\begin{center}
\includegraphics[width=0.70\textwidth]{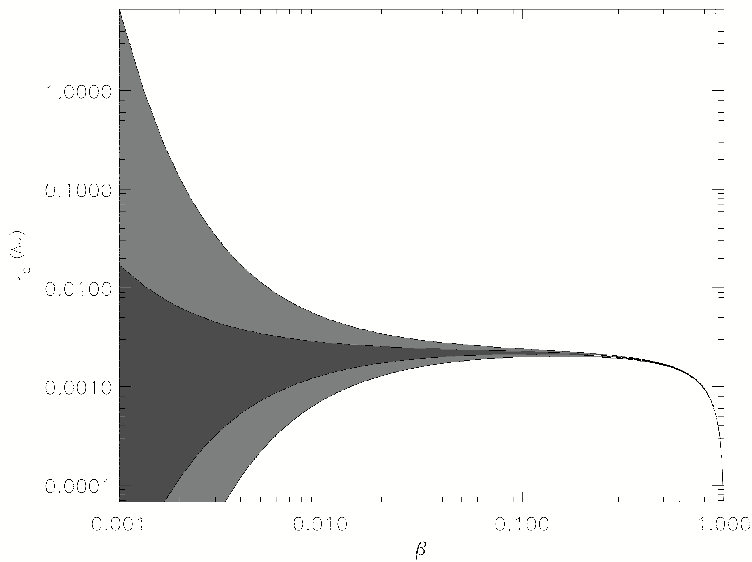}
\end{center}
%\vspace{-1.3cm}
 \caption{Constraints on the fourth order theory parameters ($\beta$ and $r_c$)
arising from the deflection angle of light rays close to the solar
limb. The grey and light grey regions embed the part of the
parameter space allowed by solar system observations at the
$2\sigma$ and $5\sigma$ confidence level, respectively. It is
noticing that for scale reason we have not plotted values of $r_c$
up to $10^4$ AU. For these cases, the observations can be
restored only for $\beta\rightarrow 0$.} %\vspace{-0.5cm}
\label{deflection}
\end{figure}

A constraint on the proposed theory can be derived by considering
the light bending effect in the Sun limb. It is well-known that in
the parameterized post-Newtonian formalism the bending angle
through which a electromagnetic light ray from a distant source is
deflected by a body with mass $m$ is  \cite{Will_2006,Will_1993}
\begin{equation}
\theta=\frac{(1+\gamma)Gm}{c^2b}(1+\cos \phi),
\end{equation}
where $b$ is the impact parameter, $\phi$ is the solar elongation
angle (between the Sun and the source as viewed from Earth) and
$\gamma$ is the post-Newtonian parameter. For GR, $\gamma=1$ and
for light rays at Sun's limb, $\theta_{\rm GR}=1.75''$. Recently,
\citet{Shapiro_2004} measured the bending angles for distant
compact sources and concluded that light bending angles follow GR
with a very high precision ($\gamma=0.9998\pm 0.0004$).
%In other words, this means that the Einstein prediction about
%bending of light rays passing in the solar vicinity is satisfied
%with a very high precision.
In other words, this means that the deflection of the light path
is well described by the GR theory. In particular, as radio
observations of distant sources have shown \citep{robertson}, the
observed and expected bending angles are related by
$\theta_{obs}=(1.001\pm 0.001) \theta_{\rm GR}$. In the framework
of the fourth order gravity theory, the deflection angle of light
rays at Sun's limb depends on both the parameters $\beta$ and
$r_c$. We explore this dependence in Fig. \ref{deflection}, by
requiring that the expected value for the bending angle is, at
least, within $2\sigma$ (grey region) or within $5\sigma$ (light
grey region) the observed value. Inspecting the same figure, it is
clear that only $\beta$-values nearby zero (corresponding to a
completely arbitrariness of $r_c$ ) are consistent with the
observed deflection angles.
%Since, as it was explained earlier, we should choose $r_c$
%parameter about 1 A.U., we can explore the range of variability
%for $\beta$ parameter. As one can see from Fig. \ref{deflection},
%which shows the deflection angle of light rays at Sun's limb as a
%function of $\beta$, only $\beta$-values nearby zero are
%consistent with deflection angles $\simeq 1.75"$.
We therefore emphasize that $\beta$-values considered in
\cite{Cap_06_04} and \cite{Cap_06_03} (i.e. $\beta =0.25, 0.43,
0.58, 0.75$) are ruled out by light deflection data.

\subsection{\label{sec:Solar_planets}Planetary Constraints}

%As stated above, we now discuss some observational consequences of
%fourth order gravity which, after fixing the core radius
%$r_c\simeq 1$ A.U., still depends on the parameter $\beta$ in the
%weak field approximation.

A stronger constraint on the fourth order gravity theory can be
obtained from the motion of the solar system planets. Let us
consider as a toy model a planet moving on circular orbit (of
radius $r$) around the Sun. From Eq.~(\ref{potential}), the planet
acceleration $a=-\partial \Phi(r)/\partial r$ is given by
\begin{equation}
a=-\frac{Gm}{2r^2}\left[1+\left(\frac{r}{r_c}\right)^{\beta}
-\beta\left(\frac{r}{r_c}\right)^{\beta}\right]~. \label{accel}
\end{equation}
Accordingly, the planetary circular velocity $v$ can be evaluated
and, in turn, the orbital period $P$ is given by
\begin{equation}
P=P_K\sqrt{2}\left[1+\left(\frac{r}{r_c}\right)^{\beta}
-\beta\left(\frac{r}{r_c}\right)^{\beta}\right]^{-1/2}~,
\label{period}
\end{equation}
where $P_K=[4\pi^2r^3/(Gm)]^{1/2}$ is the usual Keplerian period.
In order to compare the orbital period predicted by the fourth
order theory with the Solar System observations, let us define the
quantity
\begin{equation}
\frac{\Delta P}{P_K}=\frac{|P-P_K|}{P_K}=|f(\beta,r_c)-1|
\label{factor}
\end{equation}
being $f(\beta, r_c)$ the factor appearing on the right hand side
of equation (\ref{period}) and multiplying the usual Keplerian
period.

There is a question about a possibility to satisfy the planetary
period condition -- vanishing the Eq.~(\ref{factor}) -- with
$\beta$ parameter which is significantly different from zero.
Vanishing the right hand side of Eq.~(\ref{factor}) we obtain the
 relation
\begin{equation}
\ln r = \ln{r_c}-\frac{\ln{(1-\beta)}}{\beta}, \label{factor2}
\end{equation}
so that  Eq.~(\ref{factor2}) should be satisfied for all the
planetary radii. This is obviously impossible since the fourth
order theory defines $\beta$ as a parameter, while the specific
system under consideration (the Solar system in our case) allows
us  to specify the $r_c$ parameter. Hence the right hand side of
Eq.~(\ref{factor2}) is fixed for the Solar system, implying that
it is impossible to satisfy Eq.~(\ref{factor2}) even with two (or
more) different planetary radii.
\begin{figure}[t!]
\begin{center}
\includegraphics[width=\textwidth]{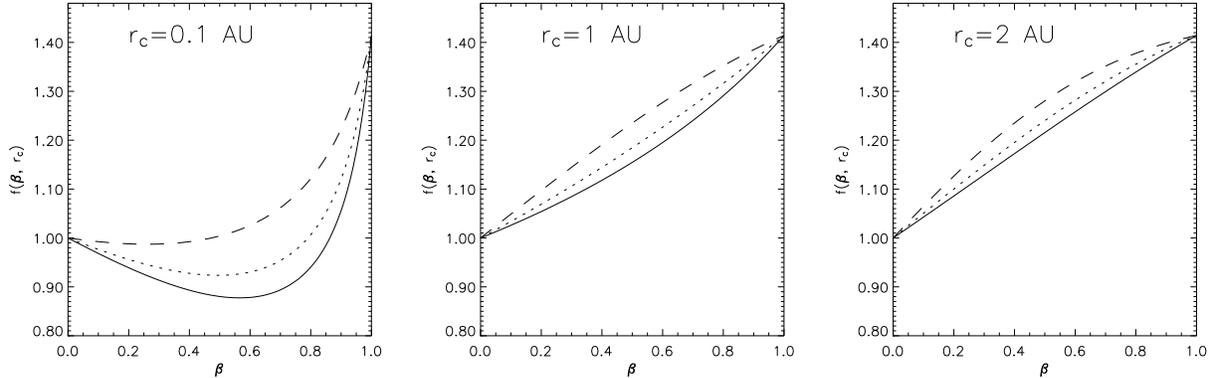}
\end{center}
%\vspace{-1.3cm}
 \caption{
The orbital period in units of the Keplerian one (the function
$f(\beta, r_c)$) is given, as a function of $\beta$, in the case
of Mercury (dashed line), Venus (dotted line), Earth (solid line)
for different $r_c$ (it is clear that $f(\beta, r_c) \rightarrow
\sqrt{2}$ for $\beta \rightarrow 1$). If $r
>r_c$ there is $\beta \in (0,1)$ satisfying Eq.~(\ref{factor2}), but
the $\beta$ values depend on fixed $r_c$ and $r$ and they are
different for a fixed $r_c$ and different $r$, so that it is
impossible to satisfy Eq.~(\ref{factor2}) if the number of planets
is more than one. Moreover, if we have at least one radius $r \leq
r_c$, there is no solution of Eq.~(\ref{factor2}).
} %\vspace{-0.5cm}
\label{planets}
\end{figure}

Just for illustration we present the function $f(\beta, r_c)$ as a
dependence on $\beta$ parameter for fixed $r_c$ and planetary
radii $r$ (see Fig. \ref{planets}). As one can see from
Eq.~(\ref{factor2}) (and Fig.~\ref{planets} as well) for each
planetary radius $r
>r_c$ there is $\beta \in (0,1)$ satisfying Eq.~(\ref{factor2}), but
the $\beta$ value depends on fixed $r_c$ and $r$, so that they are
different for a fixed $r_c$ and different $r$. Moreover, if we
have at least one radius $r \leq r_c$, there is no solution of
Eq.~(\ref{factor2}). Both cases imply that the $\beta$ parameter
should be around zero.

\begin{figure}[t!]
\begin{center}
\includegraphics[width=\textwidth]{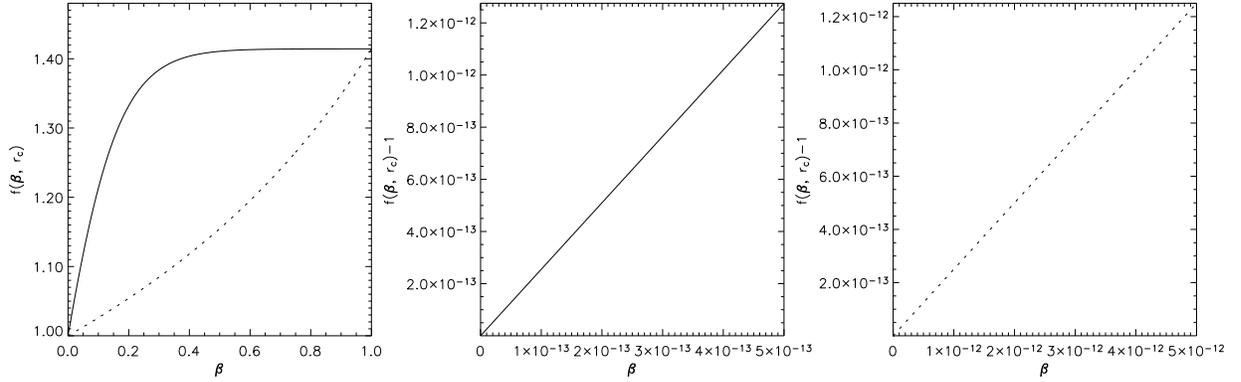}
\end{center}
%\vspace{-1.3cm}
\caption{The factor $f(\beta, r_c)$ is given as a function of
$\beta$ (left panel) for the two limiting values of $r_c$, $1$ AU
(dashed line) and $\simeq 10^4$ AU (solid line), respectively. As
one can note, only for $\beta$ approaching 0 it is expected to
recover the value of the Keplerian period. Here, the calculation
have been performed at Earth (i.e. $r= 1$ AU). In the middle and
right panel, the quantity $f(\beta, r_c)-1$ is given as a function
of $\beta$ for $r_c=10^4$ AU and $r_c=1$ AU. Note that only for
values of
$\beta$ close to $0$ the Solar System observation can be restored (see text for more details).} %\vspace{-0.5cm}
\label{figperiods}
\end{figure}

 In Fig.~\ref{figperiods} (left panel), the factor $f(\beta, r_c)$ is
given as a function of $\beta$ for the two limiting values of
$r_c$, $1$~AU (dashed line) and $\simeq 10^4$~AU (solid line),
considered in \cite{Cap_06_04}. As one can note, only for $\beta$
approaching zero it is expected to recover the value of the
Keplerian period. In the above mentioned figure, the calculation
has been performed for the Earth orbit (i.e. $r= 1$ AU).

Current observations allow also to evaluate the distances between
the Sun and the planets of the Solar System with a great accuracy.
In particular, differences in the heliocentric distances do not
exceed 10 km for Jupiter and amount to 180, 410, 1200 and 14000 km
for Saturn, Uranus, Neptune and Pluto, respectively
\citep{pietiva2005}. Errors in the semi-major axes of the inner
planets are even smaller (see e.g. Table 2 in \citep{moffat}) so
that the relative error in the orbital period determination is
extremely low. As an example, the orbital period of Earth is
$T=365.256363051$ days with an error of $\Delta T =5.0\times
10^{-10}$ days, corresponding to a relative error of $\Delta T/T$
less than $10^{-12}$. These values can be used in order to
constrain the possible values of both the parameters $\beta$ and
$r_c$ introduced by the fourth order gravity theory. This can be
done by requiring that $\Delta P/P_K\ut < \Delta T/T$ so that, in
the case of Earth, $|f(\beta, r_c)-1|\ut < 10^{-12}$ which can be
solved with respect to $\beta$ once the $r_c$ parameter has been
fixed to some value. For $r_c=1$~AU and $r_c=10^4$~AU (i.e. the
two limiting cases considered by \citet{Cap_06_04}) we find the
allowed upper limits on the $\beta$ parameter to be $4.0\times
10^{-12}$ and $3.9\times 10^{-13}$, respectively (since $\Delta
P/P_K=\Delta \beta [-1+\ln{(r/r_c)}]/4$). These results can also
be inferred from the middle and right panels of Fig.
\ref{figperiods}.

A more precise analysis which takes into account the planetary
semi-major axes and eccentricities leads to variations of at most
a few percent with respect to the results in
Fig.~\ref{figperiods}, since the planetary orbits are nearly
circular. Therefore, in spite of the fact that orbital periods of
planets are not generally used to test alternative theories of
gravity (since it is taken for granted that the weak field
approximation
 of these theories gives the Newtonian limit), we found that
these data are important to constrain parameters of the fourth
order gravity theory.

\section{\label{discussion} Discussion}

GR and Newtonian theory (as its weak field limit) were verified by
a very precise way at different scales. There are observational
data which constrain parameters of alternative theories as well.
As a result, the parameter $\beta$ of fourth order gravity should
be very close to zero (it means that the gravitational theory
should be very close to GR). In particular, the $\beta$ parameter
values considered for microlensing \cite{Cap_06_04}, for rotation
curves \cite{Cap_06_03} and cosmological SN type Ia
\cite{Borowiec_06} are ruled out by solar system data.

No doubt that one could also derive further constraints on the
fourth order gravity theory by analyzing other physical phenomena
such as Shapiro time delay, frequency shift of radio photons
\cite{Bertotti}, laser ranging for distant objects in the solar
system, deviations of trajectories of celestial bodies from
ellipses, parabolas and hyperbolas and so on. But our aim was only
to show that only $\beta\simeq 0$ values are not in contradiction
with solar system data in spite of the fact that there are a lot
of speculations to fit observational data with  $\beta$ values
significantly different from zero.

\begin{acknowledgments}
AFZ is grateful to Dipartimento di Fisica Universita di Lecce and
INFN, Sezione di Lecce where part of this work was carried out and
to the National Natural Science Foundation of China (Grant \#
10233050) and to the National Key Basic Research Foundation (Grant
\# TG 2000078404) for partial financial support. AAN, FDP and GI
have been partially supported by MIUR through PRIN 2004 - prot.
$2004020323\_ 004$. The authors are grateful to the referee whose
comments improved the paper.
\end{acknowledgments}


\end{document}